\documentclass[aps,prb,twocolumn,superscriptaddress,longbibliography]{revtex4-1}

\usepackage{hyperref} 
\usepackage{graphicx}
\usepackage{amsmath}
\usepackage{amssymb}
\usepackage{changes}
\usepackage{comment}

\DeclareGraphicsExtensions{.png,.jpg,.eps}
\usepackage{xcolor}

\begin{document}

\author{M. R\"osner}
\affiliation{Radboud University, Institute for Molecules and Materials, NL-6525 AJ Nijmegen, The Netherlands}
\email{m.roesner@science.ru.nl}

\author{J. L. Lado}
\affiliation{Department of Applied Physics, Aalto University, 00076 Aalto, Espoo, Finland}

\title{Inducing a many-body topological state of matter through
    Coulomb-engineered local interactions}

\begin{abstract}
{
	The engineering of artificial systems hosting topological excitations is at the heart of current condensed matter research.  Most of these efforts focus on single-particle properties neglecting possible engineering routes via the modifications of the fundamental many-body interactions.  Interestingly, recent experimental breakthroughs have shown that Coulomb interactions can be efficiently controlled by substrate screening engineering. Inspired by this success } we propose a simple platform in which topologically non-trivial many-body excitations emerge solely from dielectrically-engineered Coulomb interactions in an otherwise topologically trivial single-particle band structure.  Furthermore, by performing a realistic microscopic modeling of screening engineering, we demonstrate how our proposal can be realized in one-dimensional systems such as quantum-dot chains.  Our results put forward Coulomb engineering as a powerful tool to create topological excitations, with potential applications in a variety of solid-state platforms.
\end{abstract}

\date{\today}

\maketitle

\section{Introduction}

Topology represents one of the most fertile fields in modern condensed matter
physics,\cite{RevModPhys.83.1057,RevModPhys.82.3045,Ando2013} boosted by the prediction and
experimental realization of topological systems, ranging from quantum spin Hall
\cite{PhysRevLett.95.226801} and Chern\cite{PhysRevLett.61.2015,Chang2013} insulators, to topological
superconductors,\cite{Beenakker2013,Mourik2012,NadjPerge2014} and topological crystalline
insulators.\cite{PhysRevLett.106.106802,Tanaka2012}
Besides being on fundamental interest, these states are 
widely discussed due to their potential impact on solid-state 
technology, including low-consumption 
electronics,\cite{Seidel2019} spintronics\cite{mejkal2018} and 
topological quantum computing.\cite{Alicea2011} The topological
classification of non-interacting systems continues to grow
with the recent examples of higher order topological insulators,\cite{Schindler2018,Schindler2018phys}
non-Hermitian topology,\cite{PhysRevX.8.031079,PhysRevLett.120.146402,2020arXiv200801090D} 
fragile topological phases,\cite{PhysRevLett.121.126402,PhysRevX.9.021013}
quasi-periodic topology\cite{Kraus2016,PhysRevLett.109.106402}, and 
random topological systems.\cite{PhysRevLett.118.236402,Pyhnen2018}

\begin{figure}[t!]
\centering

    \includegraphics[width=\columnwidth]{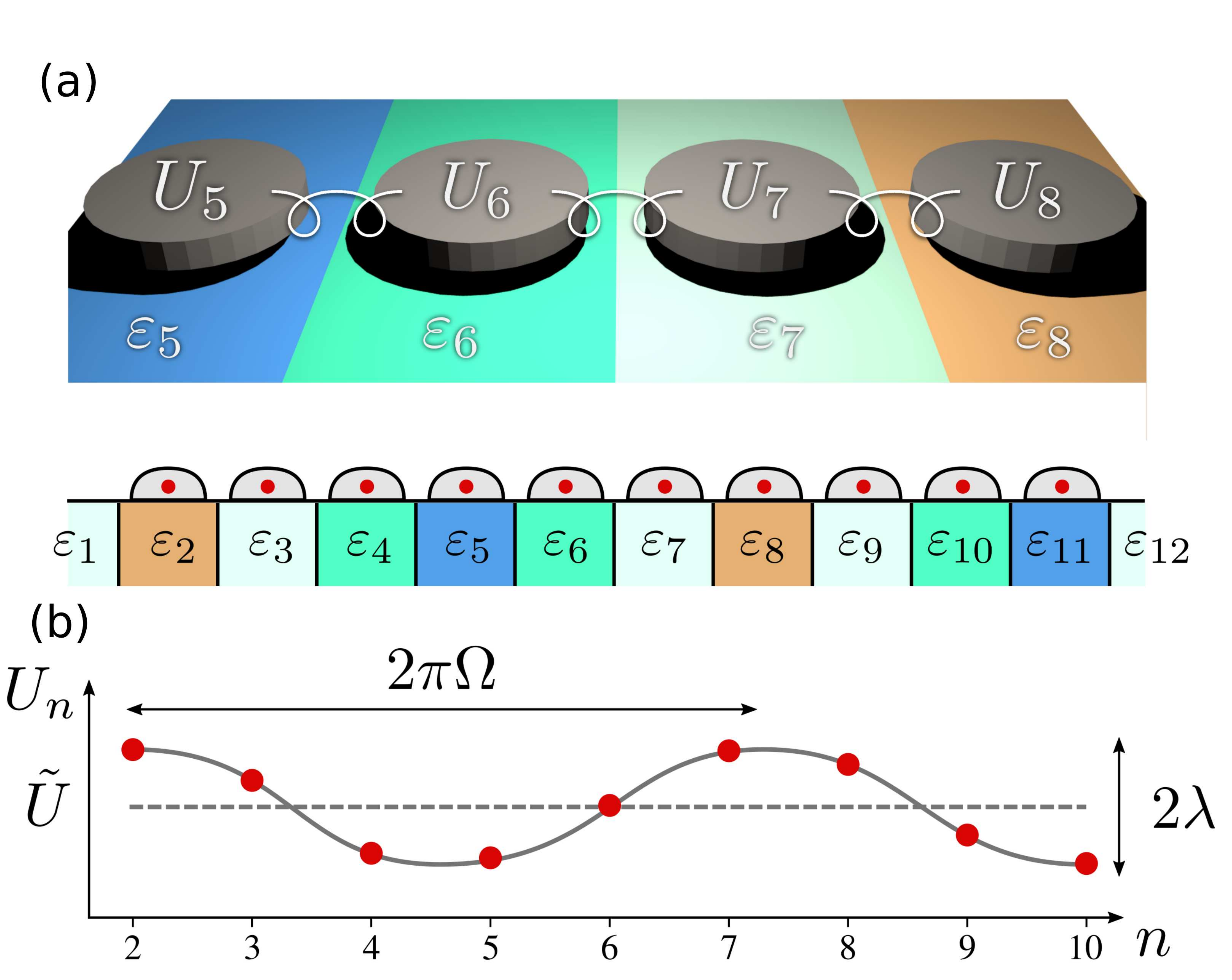}

	\caption{(a) Sketch of the proposed system: a quantum dot chain is deposited on a structured substrate.
	The spatial modulation of the substrate permitivity results in a spatially-dependent environmental screening resulting in site-dependent local Coulomb interactions (b) within the otherwise homogeneous quantum-dot chain.
}
\label{fig:fig1}
\end{figure}

Some of the most ground-breaking discoveries in condensed matter physics
have been intimately related with strong interactions, with the
paradigmatic examples of
high-temperature superconductivity\cite{RevModPhys.66.763} 
and fractional quantum Hall physics.\cite{PhysRevLett.63.199,PhysRevLett.59.1776}
It is thus not surprising that the interplay of topology and 
correlation effects is raising as one of the most enigmatic 
fields in modern condensed matter physics.
Topological states associated to interactions in topological Mott \cite{PhysRevLett.100.156401} and Kondo\cite{PhysRevLett.104.106408,Dzero2016} insulators represent first examples.
In many instances the role of interactions in generating topology
is however reduced to a mean-field 
single-particle effect.\cite{PhysRevLett.104.106408,Dzero2016,PhysRevLett.100.156401,PhysRevLett.122.016803}
The potentially genuine role of interactions in topological systems beyond
mean-field single-particle effects is hence still at an early
stage. 

Here, we demonstrate that Coulomb-engineered local interactions allow to 
induce many-body topological excitations. 
Our mechanism puts forward a paradigm to generate topological states purely driven 
by electronic interactions.
These topological states result from imprinting quasi-periodic structures to the 
local Coulomb interactions $U_n$ only, without requiring modifications to the
topologically trivial single-particle dispersions.
Crucially, we show how such a spatial structuring of the local interactions can be 
experimentally achieved using quantum-dot arrays and by exploiting Coulomb engineering\cite{rosner_two-dimensional_2016,raja_coulomb_2017,utama_dielectric-defined_2019,waldecker_rigid_2019} via structured substrate, as depicted in Figs. \ref{fig:fig1}ab. 
The manuscript is organized as follows:
in Sec. \ref{sec:single} we present how spatially modulated Coulomb interactions give rise to topological modes at the mean-field level, in
in Sec. \ref{sec:many} we show how topological modes appear from these Coulomb interactions in a purely many-body model without a single-particle analog,
in Sec. \ref{sec:coulomb} we show how
modulated Coulomb interactions can be realized by dielectric engineering,
finally in Sec. \ref{sec:con} we summarize our conclusions.

\begin{figure}[t!]
\centering
    \includegraphics[width=\columnwidth]{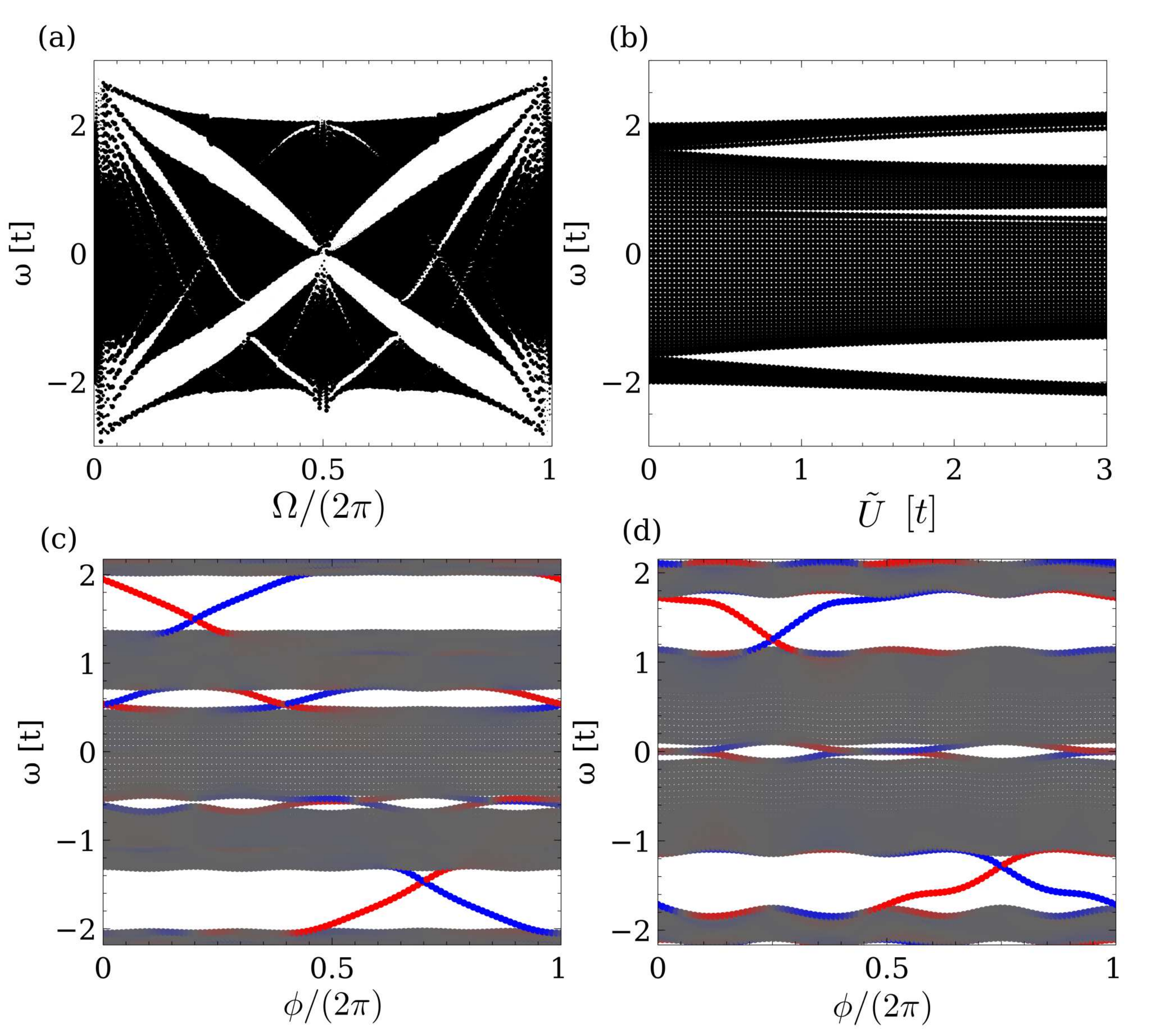}

	\caption{{\textbf{Screening-induced mean-field topology.}
	Bulk spectra as a functions of (a) $\Omega$ and (b) $\tilde{U}$ showing the emergence of topological gaps. Full 
	spectra as a function of $\phi$ for (c) $\Omega=0.4 \pi$ and (d) $\Omega=0.5\pi$ depicting edge states pumping through the gaps.
	Red/blue denotes left/right edge, the Fermi level is at $E=0.0$ and the system
	half filled. We use 80 sites and choose $\tilde{U}=6t$, $\lambda=0.5$, and $\Omega=0.4\pi$ if not stated otherwise.}
}
\label{fig:fig2}
\end{figure}

\section{Screening-Induced Single-Particle Topology}
\label{sec:single}

To exemplify our proposal {using a simple Hamiltonian}, 
we start with a model in which interactions generate a topological state
that can be understood on the 
single-particle level within a mean-field framework.
We consider a linear array of quantum dots (QDs) with a single level per dot
and spatially modulated local Coulomb interactions. The corresponding Hamiltonian reads
\begin{equation}
	H = \sum_{n,s} t \, c^\dagger_{n,s} c_{n+1,s}
	+ \sum_{n} U_n \,
	c^\dagger_{n,\uparrow}c_{n,\uparrow} 
	c^\dagger_{n,\downarrow}c_{n,\downarrow} + \text{H.c.},
	\label{eq:min}
\end{equation}
where $n$ is the QD site index, $t$ the nearest-neighbour hopping, and $U_n$ a spatially varying local Coulomb repulsion of the form
$
	U_n = \tilde{U}[1 + \lambda \cos(\Omega n + \phi) ], \label{eq:Un}
	$
which is periodically modulated around a constant $\tilde{U}$.
$\Omega$, $\phi$, and $\lambda$ are the modulation wavelength, phase, and strength, respectively, which can be efficiently controlled via spatially structured substrates as we demonstrate later.
With these parameters we are able to explore the full 
phase space of Coulomb pattern-induced topological effects. 
We solve this model via a mean field decoupling 
yielding
$
	H_{mf} = \sum_{n,s} c^\dagger_{n,s} c_{n+1,s}
	+ \sum_n U_n
	\langle c^\dagger_{n,s}c_{n,s} \rangle
	c^\dagger_{n,\bar s}c_{n,\bar s} + \text{H.c.}.
$
Assuming time reversal symmetry, i.e. $\langle c^\dagger_{n,\uparrow}c_{n,\uparrow} \rangle = \langle c^\dagger_{n,\downarrow}c_{n,\downarrow} \rangle $, the
interactions $U_n$ locally renormalize the onsite energies, following the modulation profile.
This mean-field Hamiltonian is a realization of a diagonal Aubry-Andre-Harper model\cite{Harper1955,Aubry1980AnalyticityBA} for electrons 
with a non-zero charge average per site that is known to have
edge states stemming from a parent two-dimensional Hall state\cite{PhysRevLett.49.405,PhysRevB.27.6083}.

{The topological invariant at certain excitation gaps of such models is given by the Chern number,\cite{PhysRevB.27.6083,KELLENDONK1995,PhysRevLett.49.405,2020arXiv201203644Z,PhysRevResearch.2.022049}
which match the number of states that cross the gap
as a function of the phason $\phi$.\cite{PhysRevB.27.6083,PhysRevLett.109.116404}}
In Fig.~\ref{fig:fig2} a) we show the bulk energy spectrum of a QD chain as a function of $\Omega$ at half-filling, which is clearly gapped in certain regions. 
As shown in Fig.~\ref{fig:fig2} b), those gaps 
increase with $\tilde{U}$.
These gaps are of topological origin\cite{PhysRevResearch.2.022049, zuo_topological_2020} 
as depicted in Figs.~\ref{fig:fig2} c) and d), where we show the mean-field 
spectrum as a function of $\phi$. 
In particular, we see two edge modes (red/blue) that cross the gap as $\phi$ is changed.
Spatially modulated onsite Hubbard interactions are thus indeed capable of generating
topological non-trivial modes by modulating the effective single-particle 
mean-field Hamiltonian.
This rather simple mechanism on the effective single-particle level exemplifies that interactions alone are able to induce non-trivial topology.

\begin{figure}[t!]
\centering
    \includegraphics[width=\columnwidth]{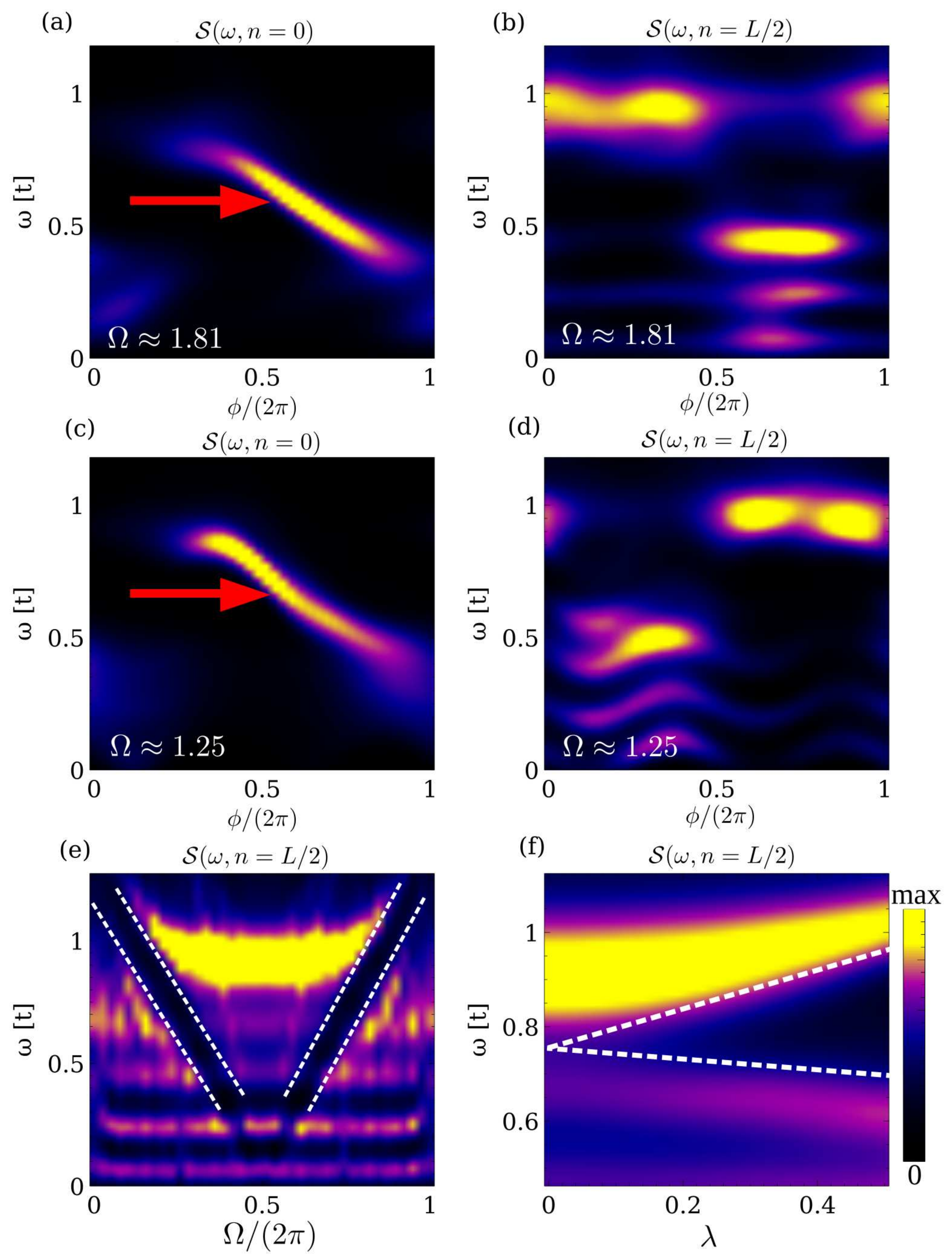}

	\caption{\textbf{Screening-induced
	many-body topology.}
	Spin spectral function in the edge (a)
	and in the bulk (b) for $\Omega=\pi/\sqrt{3}$.
	Spectral function in the edge (c)
	and in the bulk (d) for $\Omega=\pi/2$.
    (e) Bulk spectral function as a
    function of $\Omega$, showing gaps at generic
    values of $\Omega$. (f) Approx. linear scaling of the topological
    gap as a function of $\lambda$
    We used 30 sites and set $\lambda=0.5$, $\tilde{U}= 6t$, and $\Omega = 0.4 \pi$ if not stated otherwise.
}
\label{fig:fig3}
\end{figure}

\section{Screening-Induced Many-Body Topology}
\label{sec:many}
In a more sophisticated scenario we induce topological behaviour in a regime with a topologically trivial mean-field Hamiltonian.
To this end, we propose a similar interacting Hamiltonian as before but with exactly one electron per site:
\begin{eqnarray}
	H & = & \sum_{n,s} t \, c^\dagger_{n,s} c_{n+1,s} + \text{H.c.} \nonumber \\
	& & + \sum_{n} U_n
	\left (
	c^\dagger_{n,\uparrow}c_{n,\uparrow} -\frac{1}{2} 
	\right )
	\left (
	c^\dagger_{n,\downarrow}c_{n,\downarrow} -\frac{1}{2} 
	\right ). \label{eq:h2}
\end{eqnarray}
This Hamiltonian can be understood as a Hubbard QD chain
in which each QD is biased so that it is half filled.
The corresponding mean-field decoupled Hamiltonian is by construction 
uniform $H_{mf} = \sum_{n,s} t c^\dagger_{n,s} c_{n+1,s}$ and thus topologically trivial. 
To exactly treat the Hamiltonian from Eq.~\ref{eq:h2} we use the tensor
network formalism.\cite{RevModPhys.78.275,PhysRevB.91.115144,PhysRevResearch.3.013095,PhysRevResearch.2.023347,ITensor,2020arXiv200714822F,dmrgpy}
For all $\phi$ the system remains non-magnetic and half filled in every site.
We analyze the dynamical spin response defined by
$	\mathcal S (\omega,n) = \langle GS | S^z_n
	\delta (\omega - H + E_{GS})
	S^z_n |GS \rangle
$, that can be measured with inelastic spectroscopy\cite{Spinelli2014} and which is shown in Figs.~\ref{fig:fig3} a) and c) as a function of $\phi$ for different $\Omega$. Again, we find edge excitations
crossing bulk excitation gaps as the phason is changed. These finite spectral gaps are visible in the bulk spin responses shown in Figs.~\ref{fig:fig3} b) and d) and are present for arbitrary modulation frequencies $\Omega$, as depicted in Fig.~\ref{fig:fig3} e), leading to in-gap edge excitations for generic $\Omega$. These gaps are furthermore proportional to the Hubbard modulation strength $\lambda$, as shown in Fig.~\ref{fig:fig3} f). 
We verified that such in-gap modes are robust towards the presence of random disorder in the modulated local Coulomb interaction $U_n$ or inter-dot hopping $t'$.
{These characteristics underline their
topological origin\cite{PhysRevB.27.6083,PhysRevLett.109.116404,Lohse2018} which is here solely 
created by modulated many-body interactions. }

The model 
of Eq. \ref{eq:h2} is a many-body version of the Aubry-Andre-Harper model 
as introduced in the previous section. 
However, a mapping to the parent electronic two-dimensional quantum Hall state cannot be performed due to its genuine many-body nature. 
To anyway understand how the engineered interactions are
capable of creating topological edge modes here, we explore the model in the strong coupling limit, i.e. $\tilde{U} \gg t$. 
In this strongly interacting limit, spin and charge degrees of freedom are separated while every QD is still hosting one electron, giving rise to gaped charge excitations decoupled from the spin sector.
This can be explicitly shown by performing a Schrieffer-Wolff transformation of the Hamiltonian from Eq.~\ref{eq:h2} {(which is not possible for Eq.~\ref{eq:min})} into spin operators yielding the effective Hamiltonian 
\begin{equation} 
    H^{eff} = \sum_n J_{n,n+1} \bf{S}_n \cdot \bf{S}_{n+1},
\end{equation}
where $ \bf{S}_n$ are the $S=1/2$ operators on each site, and $J_{n,n+1}$ is the effective exchange interaction that takes the form 
$
    J_{n,n+1} = 2t^2 \left (\frac{1}{U_n}+\frac{1}{U_{n+1}}\right ) \approx \frac{4t^2}{\tilde{U}}[1 - \lambda \cos{(\Omega n + \phi)}]
$. \footnote{We take $\lambda \ll 1$ and $\Omega \ll 1$}
This Hamiltonian realizes a quasi-periodic 
anti-ferromagnetic $S=1/2$ Heisenberg
model,\cite{PhysRevResearch.1.033009,Agrawal2020} whose ground state is 
a time-reversal symmetric singlet state.
Such a ground state is an entangled many-body state that cannot
be described as a classical symmetry broken 
anti-ferromagnetic state due to strong quantum fluctuations.
Its low-energy excitations have $S=1/2$, in contrast to $S=1$ of classical magnets. 
A common approach to characterize these low-energy excitations is to use a 
so-called parton Abrikosov fermion transformation of the 
form $S^\alpha_n = \sum_{s,s'} \frac{1}{2} \sigma^\alpha_{s,s'} f_{n,s}^\dagger f_{n,s}$, where $f_{n,s}^\dagger$ and $f_{n,s}$ are the 
creation and annihilation spinon operators.\cite{Savary2016}
Using this to transform the operators in $H^{eff}$ followed by a mean-field decoupling for the Abrokosov fermions, we obtain an effective Hamiltonian of the form
$H^{eff}_{P} = \sum_{n,s} \gamma_{n,n+1} f_{n,s}^\dagger f_{n+1,s} + \text{H.c.}$.\footnote{We assume in the mean field ansatz that time reversal symmetry is not broken and that no anomalous terms are generated, as will be expected for the uniform Heisenberg model.} 
This effective Hamiltonian $H^{eff}_{P}$ describes 
fractionalized particles with $S=1/2$ and no charge, where the 
effective hoppings are proportional to the exchange coupling 
of the parent Heisenberg model, i.e. $\gamma_{n,n+1} \sim J_{n,n+1}$.
$H^{eff}_{P}$ thus again resembles an off-diagonal spinon
Aubry-Andre-Harper model, that can 
be mapped to a two-dimensional quantum Hall state for spinons.
As a result, the effective Hamiltonian $H^{eff}$ hosts {topologically} non-trivial 
edge {spin} excitations, and so does the original Hamiltonian from Eq.~\ref{eq:h2} in the strong-coupling limit, 
resulting from the spatially patterned local interactions.
{
The topological behaviour of the model from  Eq.~\ref{eq:h2} is thus clearly induced by many-body effects alone, 
and does not rely on any topological behaviour on the single-particle / mean-field level.
}

Let us now briefly comment on the role of edge perturbations.
A perturbation on the edge
can change the energy of the edge mode discussed above,
similarly as in 
single-particle
Su-Schrieffer-Heeger (SSH) model,\cite{PhysRevLett.42.1698,Drost2017,Grning2018}
in second order topological insulators\cite{PhysRevB.96.245115,Schindler2018,Benalcazar2017,Kempkes2019},
as well as in topological crystalline insulators.\cite{PhysRevLett.106.106802,PhysRevLett.124.236404}
In fact the topologically non-trivial states of the SSH and higher-order
topological models can be rationalized
as highly-specific examples of a generalized single-particle quasiperiodic topological
mode\cite{PhysRevResearch.2.022049}. In this regard, the topological
origin of the modes in our many-body proposal share analogous robustness as
the modes in the SSH\cite{PhysRevLett.42.1698,Drost2017,Grning2018} and higher-order models\cite{PhysRevB.96.245115,Schindler2018,Benalcazar2017,Kempkes2019}.
Finally, it is worth noting that, in the presence of an edge perturbation,
the spectrum of the presented many-body models will always show an edge mode
pumping the gap as the parameter $\phi$ is changed\cite{PhysRevB.27.6083,PhysRevLett.115.195303,PhysRevResearch.2.022049,2020arXiv201203644Z}.

\section{Local Coulomb Engineering} 
\label{sec:coulomb}

After establishing the {novel} concept of Coulomb-engineered topology 
in 1D systems, we now turn to the experimental feasibility. While 1D tight-binding 
like chains have been already created and studied using metallic 
nano-spheres \cite{krenn_squeezing_1999,maier_observation_2002}, quantum dots
\cite{lee_selective_2006,kunets_electron_2013}, and even single atoms
\cite{kim_toward_2018,kamlapure_engineering_2018}, spatially pattered Hubbard models have not been created yet. Thus, we will focus in the following on how a structuring of $U_n$ can be achieved using
substrate-screening effects.
\begin{figure}[t!]
\centering
    \includegraphics[width=0.95\columnwidth]{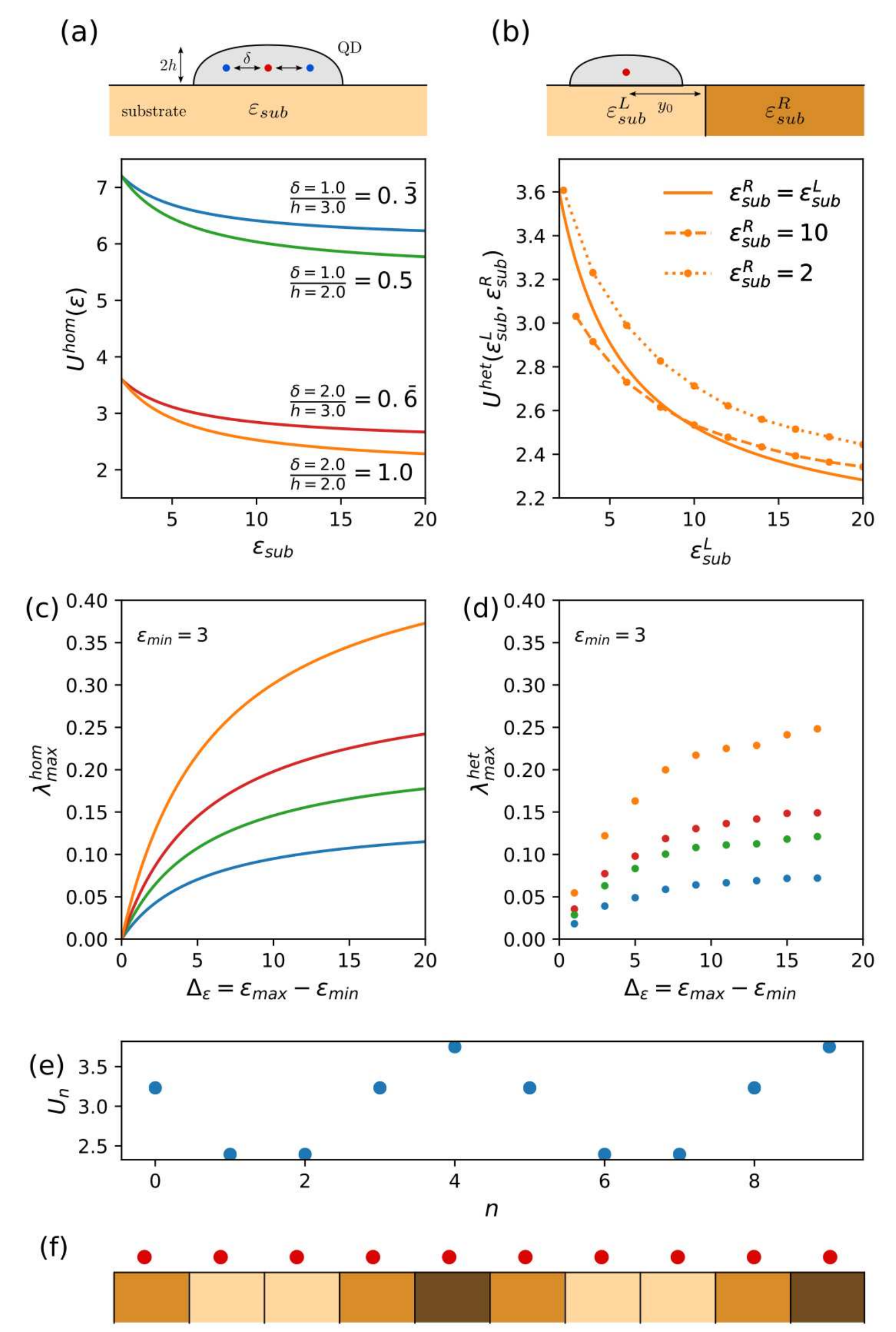}
	\caption{\textbf{Local Coulomb engineering.} Local Coulomb interaction $U_n$ controlled by (a) homogeneous and (b) heterogeneous dielectric substrates. (c)-(d) Upper and lower limits of the Coulomb modulation strength $\lambda_{max}$ estimated from homogeneous and heterogeneous substrates for different $\delta / h$ ratios and $\varepsilon_{min}=3$. (e) Example $U_n$ profile corresponding to $\tilde{U} = 3\,$eV, $\lambda=0.25$, $\Omega = \phi = 0.4\pi$ which could be realized with a patterned substrate as illustrated in the sketch (f).}
\label{fig:fig4}
\end{figure}

The local Coulomb interactions used in the Hubbard models from above are matrix elements evaluated in a \emph{single-orbital} Wannier basis $\psi(r)$ given by $U_n = \int \int dr\, dr'\, |\psi(r-r_n)|^2 |\psi(r'-r_n)|^2 \, U(r,r')$. 
In the following, we will approximate these elements by $U_n \approx e \Phi_n(\vec{\delta})$, where $\Phi_n(\vec{x})$ is the electrostatic potential of an electron with charge $e$ localized at the lattice position $n$ as felt by a second electron in its close vicinity (at distance $\vec{\delta}$), as depicted in Fig.~\ref{fig:fig4} a). 
The screening effects of a homogeneous substrate can readily be calculated using the concept of image charges, as described in the Methods section. 
$U^{hom}_n$ is than fully defined by the effective Wannier-orbital spread $\delta$ and the dot-substrate separation $h$, as depicted in Fig.~\ref{fig:fig4} a) for different $\delta$ and $h$ as a function of the substrate screening constant $\varepsilon_{sub}$ and for a background $\varepsilon_0 = 2$. 
The unscreened (bare) value of the local Coulomb interaction is
defined by $\delta$, whereas $h$ controls its vulnerability to $\varepsilon_{sub}$.
\footnote{Systems with a small Wannier-orbital spread $\delta$ have large local Coulomb
interactions, and systems with small effective heights $h$ are more strongly effected by the
substrate.} 
We see that the local Coulomb interaction can be significantly modulated  by changes to the homogeneous substrate screening.

The key element of our Coulomb-engineered topology proposal is a spatial modulation of the local interactions $U_n$. To achieve this we propose to spatially structure the substrate screening. 
A corresponding heterogeneous setup with just one dielectric interface in the substrate is depicted in Fig.~\ref{fig:fig4} b).
Here, we approximate the resulting potential $\Phi(\vec{x})$ with a multi-image-charge ansatz to fulfill the necessary boundary conditions (see Methods for more details). 
Fig.~\ref{fig:fig4} b) shows a corresponding example in form of $U^{het}(\varepsilon_{sub}^L, \varepsilon_{sub}^R)$ for $\delta/h = 1$. 
\footnote{We assume the location of the source charge above $\varepsilon_{sub}^L$ as
also depicted in the sketch of Fig.~\ref{fig:fig4} b).}  
We present data for $\varepsilon_{sub}^R = \varepsilon_{sub}^L$ (homogeneous substrate), $\varepsilon_{sub}^R =
2$, and $\varepsilon_{sub}^R = 10$, from which we see that the homogeneous solution
smoothly interpolates between the two heterogeneous situations for $\varepsilon_{sub}^L \in
[2, 10]$. 
$U^{het}(\varepsilon_{sub}^L, \varepsilon_{sub}^R = 2)$ is always the
largest due to the reduced screening from the right side of the substrate, while
$U^{het}(\varepsilon_{sub}^L, \varepsilon_{sub}^R = 10)$ is the smallest for
$\varepsilon_{sub}^L \in [2, 10]$. Most importantly, we find that there are just minor
quantitative changes to the local Coulomb interaction screened by $\varepsilon_{sub}^L$ with
$\varepsilon_{sub}^R$ being different in the close vicinity. 
We can thus conclude that periodically patterned substrate screening functions with additional dielectric interfaces will not qualitatively affect the local substrate screening properties from the immediate
surrounding.

As described above and shown in Fig.~\ref{fig:fig3}\,f), the topological gap in the spin spectral function is proportional to $t^2\lambda/\tilde{U}$. The Coulomb modulation strength $\lambda$ thus plays a significant role for our proposal as it maximizes the topological gap and thus protects the topological character of the system.
To estimate the maximal possible modulation strength, we
define
$
    \lambda_{max} = \frac{U_{max}}{U_{ave}} - 1 = \frac{\Delta_U}{U_{ave}}
$
with $\Delta_U = U_{max} - U_{min}$ and $U_{ave} = (U_{max} + U_{min})/2$. 
I.e., $\lambda_{max}$ is defined by the the maximally and minimally achievable local interactions.
For the homogeneous substrate we can calculate $\lambda^{hom}_{max}$ by defining $U_{max} = U^{hom}(\varepsilon_{min})$ and $U_{min} = U^{hom}(\varepsilon_{max})$ with $\varepsilon_{min} < \varepsilon_{max}$. 
In Fig.~\ref{fig:fig4} c) we show the resulting values for fixed $\varepsilon_{min} = 3$. 
$\lambda^{hom}_{max}$ steadily increases with the dielectric contrast $\Delta_\varepsilon$, which is driven by the enhacement of $\Delta_U$ (due to the reduction of $U_{min}$) upon increasing $\varepsilon_{max}$. 
Additionally, $\lambda^{hom}_{max}$ increases with the $\delta/h$ ratio, which results from a decreased $U_{ave}$ for increased $\delta$ and the enhanced substrate-screening vulnerability of $U^{hom}$ upon decreasing $h$. 
$\lambda^{hom}_{max}$ is thus maximized by a large dielectric
contrasts and large $\delta/h$ ratios.

These homogeneous $\lambda^{hom}_{max}$ are, however, just upper limits. 
In a more realistic setting, $U_{max}$ and $U_{min}$ might result from a heterogeneous substrate with additional dielectric interfaces as depicted in the sketch of Fig.~\ref{fig:fig4}~ef). 
To estimate $\lambda^{het}_{max}$ in such a setting we imagine the transition from $U_{max}$ to $U_{min}$ taking place within three lattice sites and set $U_{max} = U^{het}(\varepsilon_{min}, \varepsilon_{mid})$ and $U_{min} = U^{het}(\varepsilon_{max},
\varepsilon_{mid})$ with $\varepsilon_{mid} = \varepsilon_{min} + \Delta_\varepsilon / 2$. The resulting values are shown in Fig.~\ref{fig:fig4}\,d). 
Due to the additional dielectric interface in the substrate, $\Delta_U$ is decreased so that $\lambda^{het}_{max}$ is overall smaller than $\lambda^{hom}_{max}$, but behaves otherwise similar. 
For $\delta = 2$\,\AA\ (orange and red dots) $\lambda^{het}_{max}$ represents the lower limit since the interface is positioned here at $y_0 = \delta = 2\,$\AA. By increasing $y_0$ $\lambda^{het}_{max}$ approaches the upper limit $\lambda^{hom}_{max}$.
The optimal parameter regime to realize screening-induced many-body topology is thus defined by small QD heights $h$, large QD diameters $\delta$, large QD separations $y_0$, and large dielectric contrasts $\Delta_\varepsilon$ which increases $\lambda_{max}$ and thus the topological gap.

\section{Conclusions} 
\label{sec:con}

We proposed a new family of topological states,  whose  topological  excitations 
stem  purely from many-body interactions instead from single-particle engineering.
Our proposal compares with conventional schemes
that rely on engineering single-particle physics, 
demonstrating that engineered electronic interactions are a 
powerful complementary tool for exploring novel quantum states{, which can be also used to create more complex models such as
second order topological insulators.\cite{Schindler2018,Benalcazar2017,PhysRevResearch.2.022049}}
{Importantly, we showed that Coulomb-engineering can be efficiently utilized
to create the needed interaction profiles, demonstrating
that our proposal can be implemented with current dielectric-engineering
techniques.}
Our results thus put forward a new method to create topological states of matter based on
engineered interactions, providing a stepping stone to exploit dielectric engineering to
realize exotic quantum excitations.

\textbf{Acknowledgements:}
We thank L. Muechler for fruitful discussions.
J.L.L. acknowledges the computational resources provided by
the Aalto Science-IT project,
and financial support from the
Academy of Finland Projects No.
331342 and No. 336243.

\appendix

\section*{Appendix}

\section{Tensor Network formalism}
We exactly solve the many-body problems from the main text with the help of the kernel polynomial tensor network formalism.\cite{RevModPhys.78.275,PhysRevB.91.115144,PhysRevResearch.2.023347,PhysRevResearch.3.013095,ITensor,2020arXiv200714822F,dmrgpy}
Within the latter we expand the spectral function in 
terms of Chebyshev polynomials, whose coefficients can be efficiently computed using a recursion relation between tensor network wave functions.
$\mathcal{S}(n,\omega)$ is subsequently represented in terms of $N$ Chebyshev polynomials
$T_k(\omega)$ as
$
\mathcal{S}(\omega,n)=
        \frac{1}{\pi\sqrt{1-\omega^{2}}}
        \left(\mu_{0}+2\sum_{k=1}^{N}\mu_{k}T_{k}(\omega)\right)
$. The coefficients $\mu_k$ are defined by $\mu_{k}=\langle GS| S^z_n T_{k}(H) S^z_n| GS \rangle$,
which can be efficiently computed using the Chebyshev recursion relations.
Once the first $N/2$ moments are computed, we use an autoregressive algorithm\cite{Akaike1969}
to predict the next $N/2$ moments and reconstruct the spectral function using a Jackson
kernel.\cite{Jackson1912} 
The autoregressive model halve the calculation costs and the Jackson kernel quenches Gibbs oscillations.
This formalism allows to compute dynamical response functions of the many-body system
directly in frequency space, 
without requiring any time evolution.

\begin{figure}[t!]
\centering
    \includegraphics[width=\columnwidth]{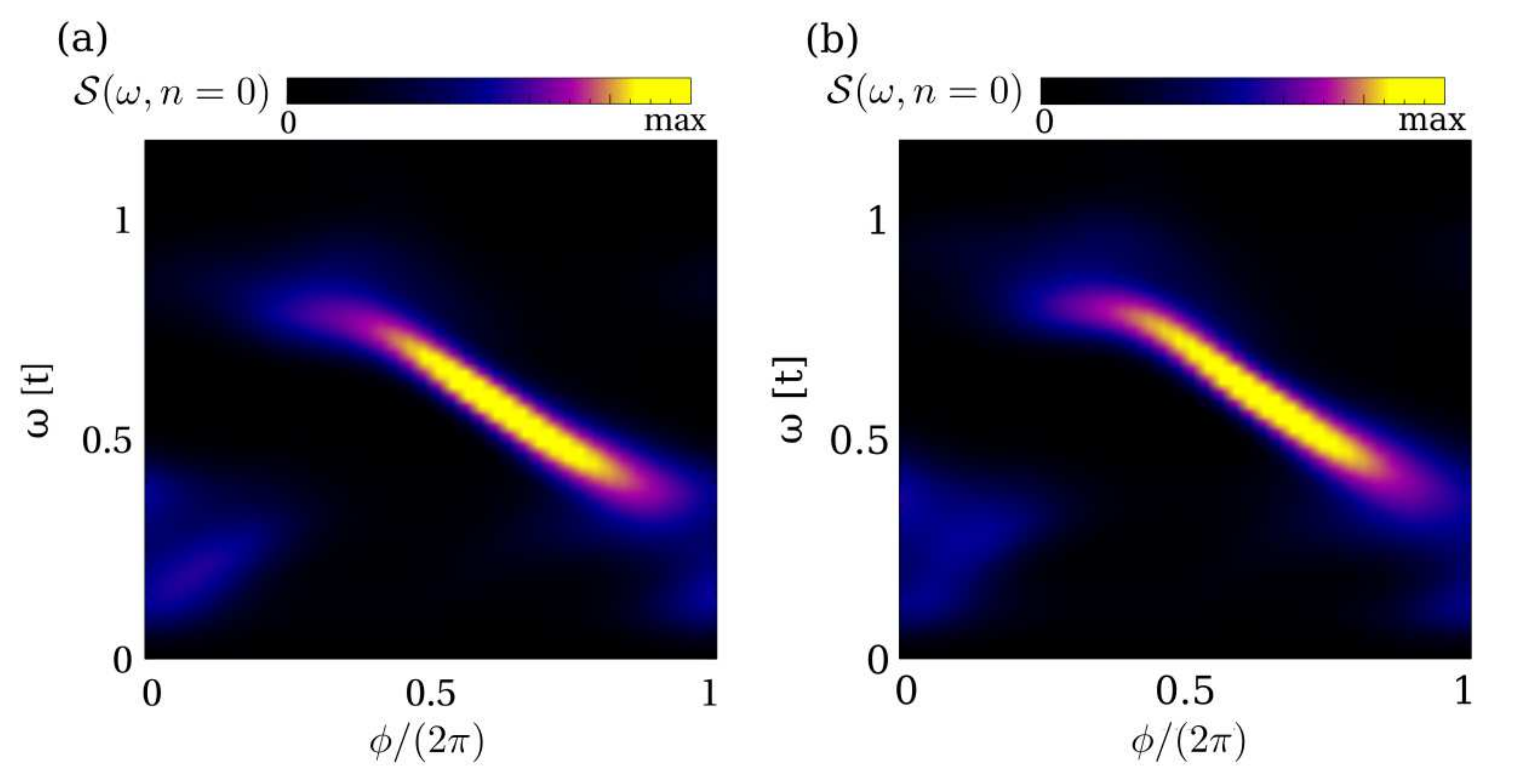}
	\caption{\textbf{Robustness of the edge modes:}
	 a) Spectral function
	of the edge for a finite second neighbor hopping, and b) with a finite
	random disorder in the local interactions, showing that the
	topological edge modes survive those perturbations.
	We took $\tilde U = 6t$ and $\Omega = \pi/\sqrt{3}$.
}
\label{fig:SMfig1}
\end{figure}

\section{Robustness of the Edge Modes} 
Here we show how the topological
edge modes are found to be robust against perturbations
of the many-body Hamiltonian of Eq. \ref{eq:h2}.
In particular, we explore two different perturbations that are especially relevant for the experimental realization:
second neighbor hopping and disorder in the interactions. 
First, the nearest neighbor hopping model of Eq. \ref{eq:h2} is expected to be an approximation to the real system,
as a finite overlap between second neighbor sites is expected. 
We capture this by adding a perturbation of the form $H_{NNN}= t_{NNN}\sum_{n,s} c^\dagger_{i,s} c_{i+2,s} + h.c.$. 
In Fig.~\ref{fig:SMfig1}~a) we show the spectral function at the edge under the influence of this additional perturbation for $t_{NNN}=0.1t$, showing that the in-gap modes survive.
Second, as our proposal requires to engineer different dielectric environments for each dot, defects in the fabrication are expected giving rise to imperfect interaction profiles. 
This can be captured by adding to the Hamiltonian from Eq.~\ref{eq:h2} a term of the form
$H_W= \sum_{n} W_n
	\left (
	c^\dagger_{n,\uparrow}c_{n,\uparrow} -\frac{1}{2} 
	\right )
	\left (
	c^\dagger_{n,\downarrow}c_{n,\downarrow} -\frac{1}{2} 
	\right )$ where $W_n$ are site-dependent random numbers.
	We show in Fig.~\ref{fig:SMfig1}~b) the spectral
function at the edge including this onsite Coulomb disorder for $W_n\in(-0.3t,0.3t)$, showing that the in-gap edge modes again survive disorder. 
These results highlight the robustness of the edge modes towards perturbations that are likely to present in the experimental setup.

\begin{figure*}[t!]
\centering
    \includegraphics[width=2.0\columnwidth]{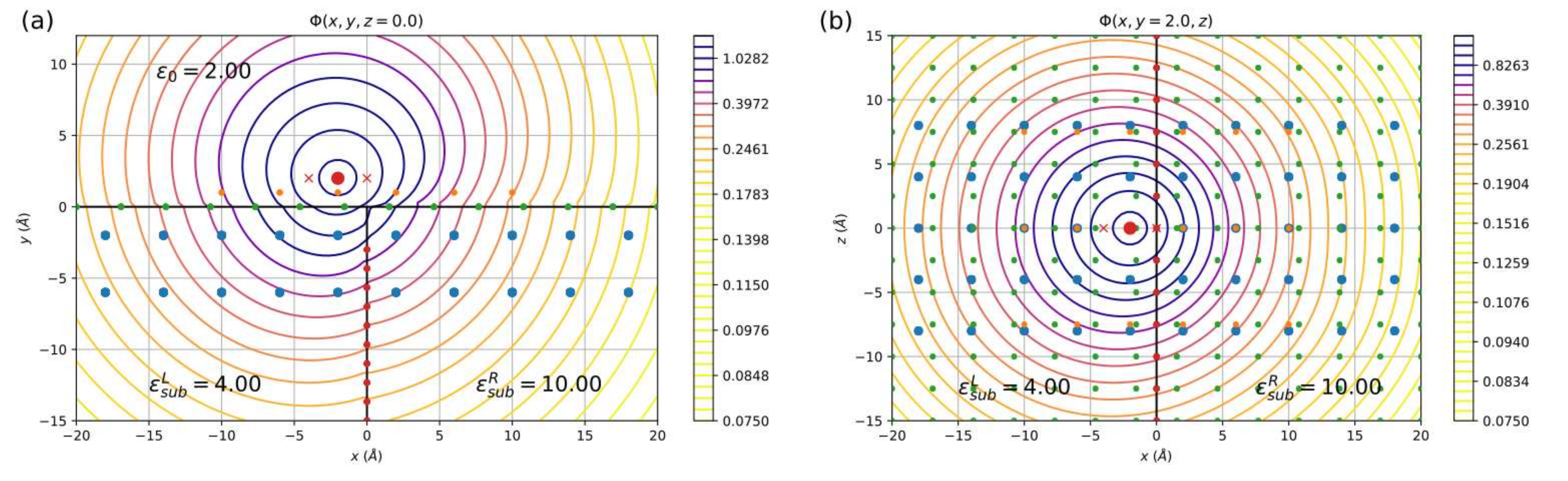}
	\caption{ Example Result of the Multi-Image-Charge Approach. Left (a) and right (b) show the numerically evaluated potential for $\varepsilon_0 = 2$, $\varepsilon_{sub}^L = 4$, $\varepsilon_{sub}^R = 10$, $h = 2\,$\AA, and $\delta=y_0 = 2\,$\AA\, in the $(x,y)$ and $(x,z)$ planes. Red dot: Source charge position, orange dots: Poisson equation evaluation points, green dots: $y=0$ interface discrete boundary condition points, red dots: $x=0$ interface discrete boundary condition points, blue dots: image charge positions, black lines: dielectric interfaces. }
\label{fig:fig5}
\end{figure*}
\section{Poisson Solver} 
To calculate the substrate-screened local Coulomb interactions $U_n = e\Phi(\delta)$ we use the concept of image charges. In the case of a simple homogeneous substrate, we can choose an ansatz of the form
\begin{align}
    \Phi(\vec{x}) = \left\{
        \begin{array}{lll}
             \frac{1}{\varepsilon_0} \frac{q_{00}}{|\vec{q}_{00} - \vec{x}|} \, +
             &\frac{1}{\varepsilon_0} \frac{q_{01}}{|\vec{q}_{01} - \vec{x}|} 
             & \text{ for } y \geq 0 \\
             &\frac{1}{\varepsilon_{sub}} \frac{q_{11}}{|\vec{q}_{11}-\vec{x}|}  
             & \text{ for } y < 0
        \end{array}
        \right.,
\end{align}
where $q_{00}$ and $\vec{q}_{00}$ are the charge and position of the source charge, and $q_{01}$, $q_{11}$, $\vec{q}_{01}$, and $\vec{q}_{11}$ are the charges and positions of the image charges. This potential needs to solve the Poisson equation in each $\varepsilon_i$ region, i.e.
\begin{align}
    \Delta \Phi(\vec{x}) = \left\{
        \begin{array}{cl}
             \frac{\rho(\vec{x})}{\varepsilon_0}
             & \text{ for } y \geq 0 \\
             0 
             & \text{ for } y < 0
        \end{array}
        \right., \label{eq:Poisson}
\end{align}
where $\rho(\vec{x})$ is the point-charge density of the source charge $q_{00}$, and must full fill the boundary conditions
\begin{align}
    \Phi(\vec{x}_{y^+}) &= \Phi(\vec{x}_{y^-}) \\
    \varepsilon_0 \frac{\partial \Phi(\vec{x}_{y^+})}{\partial y} &= 
    \varepsilon_{sub} \frac{\partial \Phi(\vec{x}_{y^-})}{\partial y}
\end{align}
at the dielectric interface defined by $y = 0$. By exploiting the full rotational symmetry around the $y$-axis (through the source charge), we can readily fix  $\vec{q}_{01}$ to the $y=0$-plane mirrored position of $\vec{q}_{00}$ and choose $\vec{q}_{11} = \vec{q}_{00}$. Subsequently $q_{01}$ and $q_{11}$ are fixed by the boundary conditions yielding 
\begin{align}
    q_{01} &= \frac{\varepsilon_0 - \varepsilon_{sub}}{\varepsilon_0 + \varepsilon_{sub}}  \, q_{00}, \\
    q_{11} &= q_{01} - q_{00}.
\end{align}
Thus, in the case of a homogeneous substrate the influence of $\varepsilon_{sub}$ to the local Coulomb interaction $U_n$ can be calculated analytically. As soon as there is an additional dielectric interface in the substrate, we cannot find an analytic solution any more. In order to estimate the impact of this heterogeneous substrate screening, we construct an approximate solution from a multiple image charge ansatz of the form
\begin{align}
    \Phi(\vec{x}) = \left\{
	    \begin{array}{lll}
             \frac{1}{\varepsilon_0} \frac{q_{00}}{|\vec{q}_{00} - \vec{x}|} \, +
             &\frac{1}{\varepsilon_0} \sum_i^{2N} \frac{q_{0i}}{|\vec{q}_{0i} - \vec{x}|} 
             & \text{, } y \geq 0 \\
             &\frac{1}{\varepsilon_{sub}^L} \sum_i^{N} \frac{q_{1i}}{|\vec{q}_{1i}-\vec{x}|}  
             & \text{, } y < 0, x \leq 0
             \\
             &\frac{1}{\varepsilon_{sub}^R} \sum_i^{N} \frac{q_{2i}}{|\vec{q}_{2i}-\vec{x}|}  
             & \text{, } y < 0, x > 0
        \end{array}
        \right.,
\end{align}
which is supposed to solve the Poisson equation given in Eq.~(\ref{eq:Poisson}) with the boundary conditions at the $y = 0$ interfaces
\begin{align}
    \Phi(\vec{x}_{y^+}) &= \Phi(\vec{x}_{y^-}) && y = 0 \label{eq:PoissonBc1} \\
    \varepsilon_0 \frac{\partial \Phi(\vec{x}_{y^+})}{\partial y} &= 
    \varepsilon_{sub}^L \frac{\partial \Phi(\vec{x}_{y^-})}{\partial y} && y = 0, x \leq 0 \\
    \varepsilon_0 \frac{\partial \Phi(\vec{x}_{y^+})}{\partial y} &=
    \varepsilon_{sub}^R \frac{\partial \Phi(\vec{x}_{y^-})}{\partial y} && y = 0, x > 0
\end{align} 
and at the $x=0$ interface in the substrate
\begin{align}
    \Phi(\vec{x}_{x^+}) &= \Phi(\vec{x}_{x^-}) && y < 0, x = 0  \\
    \varepsilon_{sub}^L \frac{\partial \Phi(\vec{x}_{x^+})}{\partial x} &=
    \varepsilon_{sub}^R \frac{\partial \Phi(\vec{x}_{x^-})}{\partial x} && y < 0, x = 0. \label{eq:PoissonBc5}
\end{align}

If we fix all image charge positions we can use the $4N$ image charges $q_{(0,1,2)i}$ to fulfill $4N$ boundary conditions at discrete interface positions $\vec{x}_i$. The resulting linear equation system is well defined and has a unique solution. Here, however, we reformulate the Poisson equation into a minimization problem of the form 
\begin{align}
    \operatorname{min}_{q_{(0,1,2)i}} \left\|
        \begin{array}{cl}
             \Delta \Phi(\vec{x}) - \frac{\rho(\vec{x})}{\varepsilon_0}
             & \text{ for } y \geq 0.0 \\
             \Delta \Phi(\vec{x}) 
             & \text{ for } y < 0.0
        \end{array}
        \right\|, 
\end{align}
using the boundary conditions from Eqs.~(\ref{eq:PoissonBc1}-\ref{eq:PoissonBc5}) to define constraints for the minimization. This relaxes the one-to-one correspondence between the number of image charges and the number of discrete boundary conditions.

Numerically, we use the Sequential Least SQuares Programming (SLSQP) algorithm \cite{kraft_software_1988} as implemented in the Scipy minimization package. We use $5 \times 2 \times 5$ image charges per substrate region (i.e. $100$ in total), separated by $(d_x, d_y, d_z) = (4, 4, 4)\,$\AA\ and distributed as indicated in Fig.~\ref{fig:fig5}, and $182$ boundary condition points at the $y=0$ interface and $130$ points at the $x=0$ interface distributed as also indicated in Fig.~\ref{fig:fig5}. To minimize the function defined in Eq.~(\ref{eq:min}) we evaluate the Poisson equation on a discrete grid of $18$ points in the $x/z$ plane slightly below the source charge (see Fig.~\ref{fig:fig5}). The source charge is positioned at $\vec{q}_{00} = (h, y_0 = -2, 0)$ (with variable distance $h$ to the $y=0$ interface and fixed position relative to the vertical interface in the substrate) and its potential is approximated using a three-dimensional Gauss function of the form
\begin{align}
    \rho(\vec{x}) = \frac{q_{00}}{\sigma^3 (2\pi)^{3/2}} \operatorname{exp}\left( - \frac{(\vec{q}_{00} - \vec{x})^2}{\sigma^2} \right)
\end{align}
with $\sigma = 0.15\,$\AA. Finally, $U_n = e \Phi(\vec{\delta})$ is evaluated using $eq_{00} \approx 14.39\,$eV\AA\, and $\vec{\delta} = \vec{q}_{00} + (\delta, 0, 0)$ with variable $\delta$. 

In Fig.~\ref{fig:fig5} we show a corresponding example result for $\varepsilon_0 = 2$, $\varepsilon_{sub}^L = 4$, $\varepsilon_{sub}^R = 10$, $h = 2\,$\AA\,, and $\delta=|y_0|=2\,$\AA. In this situation we would a priori expect to see equipotential lines with small kinks at the $\varepsilon_0$/$\varepsilon_{sub}^L$ boundary and with enhanced kinks at the interfaces to the $\varepsilon_{sub}^R$ area, which is approximately the case in our numerical estimate. Furthermore, we see that these boundary conditions are also approximately full filled between the discrete boundary positions in the vicinity of the source charge (red dot). Further away, we, however, also find deviations to this. We thus expect our numerical solution to by approximately valid in the vicinity of the source charge, which is enough to estimate the local Coulomb interaction.

\bibliography{biblio}{}

\end{document}